\documentclass[%
 reprint,
superscriptaddress,
 aps,
prl,
]{revtex4-2}

\usepackage{graphicx}
\usepackage{dcolumn}
\usepackage{bm}

\begin{document}

\preprint{APS/123-QED}

\title{Light-induced percolative topological phase transition\\ in type-II Weyl semimetal WTe$_2$}

\author{Xiaoyue Zhou}
\thanks{The authors contributed equally to this work}%
\author{Fu Deng}
\thanks{The authors contributed equally to this work}%
\author{Yifan Gao}
\author{Yi Chan}
\affiliation{%
Department of Physics, Hong Kong University of Science and Technology, Clear Water Bay, Kowloon, Hong Kong, China}%

\author{Shulei Li}
\affiliation{%
School of Optoelectronic Engineering, Guangdong Polytechnic Normal University, Guangzhou, China}%

\author{Ning Wang}
\affiliation{%
Department of Physics, Hong Kong University of Science and Technology, Clear Water Bay, Kowloon, Hong Kong, China}%
\affiliation{%
William Mong Institute of Nano Science and Technology, Hong Kong University of Science and Technology, Kowloon, Hong Kong, China}

\author{Junwei Liu}
\affiliation{%
Department of Physics, Hong Kong University of Science and Technology, Clear Water Bay, Kowloon, Hong Kong, China}%

\author{Jingdi Zhang}
\email{jdzhang@ust.hk}
\affiliation{%
Department of Physics, Hong Kong University of Science and Technology, Clear Water Bay, Kowloon, Hong Kong, China}%
\affiliation{%
William Mong Institute of Nano Science and Technology, Hong Kong University of Science and Technology, Kowloon, Hong Kong, China}

\date{\today}

\begin{abstract}
We report on an ultrafast terahertz free-carrier dynamic study of a photo-excited \textit{T$_d$}-WTe$_2$ thin film. In the photo-excited state, we observe a metastable electronic state featuring negative differential terahertz photoconductivity and reduced scattering rate. Detailed electrodynamics analysis and first-principal calculation attribute it to light-induced topological phase transition, reducing density of states near the Fermi level. Furthermore, the emergence of an unconventional temporal isosbestic point marks a dynamic universality, strongly suggesting a percolative interaction between the two topologically distinct phases. 

\end{abstract}
\maketitle

The Weyl fermion, featuring topologically protected band crossing points of linear dispersion in the momentum space, has recently been identified in various semimetals of broken inversion or time-reversal symmetry. Tungsten ditelluride (WTe$_2$) in the equilibrium \textit{T$_d$} phase of broken inversion symmetry has been predicted by theory \cite{soluyanov2015type4} and verified by experiment \cite{wang2016observation5,sanchez2016surface6} to accommodate pairs of Weyl nodes with opposite chirality. The distinct tilted conical dispersion makes it an archetypal type-II Weyl fermion system, in which Lorentz invariance is absent. The bulk electronic state of WTe$_2$ is found to be sensitive to changes in interlayer interaction \cite{qi2022traversing11} as well as correlation and renormalization effect \cite{di2017three22,beaulieu2021ultrafast23,pan2015pressure24}, laying the ground for realizing the structural and electronic control over topological phase transition \cite{basov2017towards12}. For instance, various techniques such as charge injection, orbital-selective photoexcitation and external pressure have been employed to drive transition in WTe$_2$ from the non-centrosymmetric \textit{T$_d$} phase towards the centrosymmetric 1\textit{T'} phase \cite{guan2021manipulating13,kim2017origins14,zhou2016pressure15,rossi2020two16} for a potential gateway to tune the Weyl nodes.

Notably, such structural-driven topological transition in WTe$_2$ can also be triggered by a dynamically restored inversion symmetry upon ultrafast laser excitation, which strongly modulates the interlayer interaction by effect of coherent shear phonon or photo-doping or both \cite{qi2022traversing11,ji2021manipulation17,hein2020mode18,sie2019ultrafast19,soranzio2022strong20,zhang2019light21}. Although many expect the photoexcitation to qualitatively alter the electronic band structure \cite{beaulieu2021ultrafast23}, the microscopic dynamics responsible for the transition is still poorly understood and awaits confirmation from direct spectroscopic tools, particularly those of superb spectral resolution near the Fermi level. Such dynamical insight may be accessed by techniques that track electronic and vibrational aspects of the transition on a fundamental time and spatial scale. In this work, we utilize ultrafast optical pump-THz probe spectroscopy (OPTP) probing low-energy electrodynamics of free carriers to capture the presumed band reconstruction process during the light-induced \textit{T$_d$} - 1\textit{T'} transition, assured by its high temporal- and energy-resolution. More importantly, our results identified a time-domain version of isosbestic point, suggestive of the percolative nature of the light-induced topological phase transition.

\begin{figure}
\includegraphics{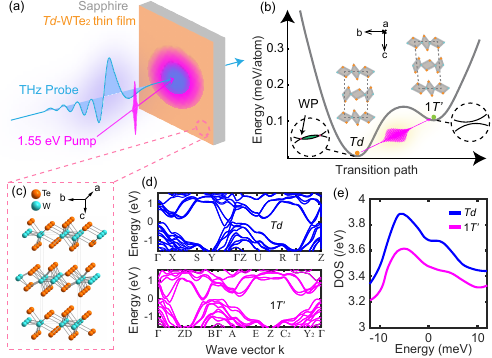}
\caption{
(a) Schematic of the optical pump-THz probe experiment. (b) DFT calculation on energies of the competing topological nontrivial \textit{T$_d$} and trivial 1\textit{T'} phases. A transition may be driven by pulsed photoexcitation. WP: Weyl point. (c) The lattice structure of \textit{T$_d$}-WTe$_2$. (d) DFT calculation on the band structure of the distinct phases along the high symmetry direction. (e) Calculated DOS near the Fermi level.
}
\label{fig1}
\end{figure}

Schematically shown in Figure \ref{fig1}(a), a near-IR pump pulse (100 fs in duration, 1.55 eV in photon energy) and a THz probe pulse with a variable delay are employed in the OPTP experiment. The 30 nm WTe$_2$ thin film is grown on a sapphire plate and in the topological nontrivial \textit{T$_d$}-phase of an orthorhombic layered lattice structure (\textit{Pmn2$_1$} space group). As shown in Fig. \ref{fig1}(c), within the ab-plane, atoms are covalently bonded with a broken inversion symmetry along the a-axis. In the out-of-plane direction, neighboring WTe$_2$ layers are weakly coupled by the Van Der Waals force, and the pertinent interlayer antibonding states were found to constitute the low-energy electronic states in the vicinity of the Fermi level \cite{kim2017origins14}. The corresponding band structure of \textit{T$_d$}-WTe$_2$ from density functional theory (DFT) calculation is shown in Fig. \ref{fig1}(d). The hole and electron pockets in close proximity to the Fermi level along the Y-$\Gamma$ direction are consistent with the results from photoemission experiments \cite{pletikosic2014electronic25,jiang2015signature26}. On the other hand, WTe$_2$ in the topological trivial 1\textit{T'} phase shows a slightly different crystalline structure (\textit{P2$_1$/m} space group) with the primary indicator being the interlayer shear displacement (Fig. \ref{fig1}b). This results in only a minor difference in free energy between the two phases, which makes feasible the transition into the 1\textit{T'} phase when certain conditions are fulfilled. As revealed by DFT calculation, 1\textit{T'}-WTe$_2$ possesses similar electron and hole pockets but with distinctions (Fig. \ref{fig1}d). The \textit{T$_d$} - 1\textit{T'} structural transition can bring about a tiny decrease in the density of states (DOS) within a minute energy scale near the Fermi level (Fig. \ref{fig1}e), which can serve as an indicator of the transition and be best resolved by ultrafast THz spectroscopy.

\begin{figure}
\includegraphics{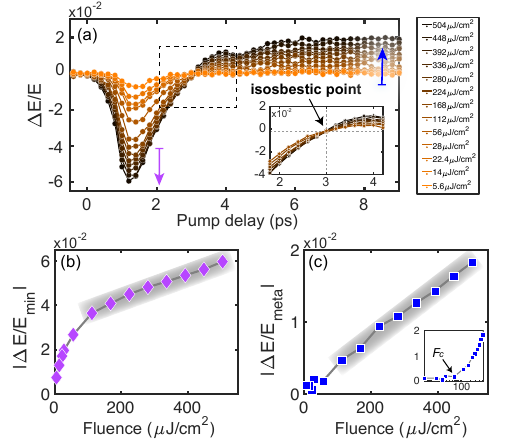}
\caption{
(a) Pump-induced differential change in peak of transmitted THz field at various fluences. The inset shows the appearance of a dynamic isosbestic point at above-threshold fluences. (b-c) Fluence dependence of the picosecond negative transient and long-lived positive offset. The inset in panel (c) shows on logarithmic scale the threshold fluence for access to the metastable state. The grey shade highlights the anomalous regime of linear scaling.
}
\label{fig2}
\end{figure}

We first employ a standard setup to study the ultrafast free-carrier response of WTe$_2$ to pulsed near-IR excitation by tracking the photo-induced differential change in peak amplitude of the transmitted THz probe pulse ($\Delta$E/E) as a function of pump-probe delay \textit{t$_p$}. Due to the negligible temperature dependence of equilibrium \textit{T$_d$} phase, we chose a base temperature of 80 K for all measurements to minimize the effect from the thermally excited carriers. Overall, the $\Delta$E/E signal follows a bimodal dynamics and its dependence on excitation fluence is shown in Fig. \ref{fig2}(a). Within the first picosecond (\textit{t$_p$} $<$ 1 ps), the negative transient in $\Delta$E/E represents an increasingly strong absorption of THz field, resulting from creation of nonequilibrium distribution of carriers at energy levels remote from equilibrium-state Fermi level. In a few hundred femtoseconds after the excitation, the photo-injected carriers of excessive energy quickly thermalize through carrier-carrier scattering and, in turn, regenerate even more carriers. At roughly a delay of 1 ps, this sudden change in electronic occupancy leads to a transient state of maximum absorption to the THz field, indicating the largest amplitude in intra-band transition. For the early stage post-transient dynamics (1 ps $<$ \textit{t$_p$} $<$ 3 ps), it follows a fast decay dynamics, corresponding to the cooling process dominated by scattering between hot carriers and a reservoir of cold phonons \cite{zhang2018ultrafast29} until reaching a quasi-equilibrium state with an occupancy akin to the that in the static state. Conventionally, the $\Delta$E/E signal becomes vanishingly small, as observed in most other gapless semimetal materials\cite{lu2018terahertz27,george2008ultrafast28}. Nevertheless, from near completion of its decay and onward (\textit{t$_p$} $>$ 3 ps), the signal surprisingly saturates to a \textit{finite} and \textit{positive} offset; a photo-induced transparency which we attribute to a reduced DOS by the reconstructed band dispersion at the Fermi level. Compared to dynamic study of the light-induced metastable phase in \textit{T$_d$}-WTe$_2$ by other probing methods \cite{ji2021manipulation17,hein2020mode18,sie2019ultrafast19}, the timescale observed by THz spectroscopy for the anomalous long-lived positive offset (photo-induced transparency) to emerge is the same and, therefore, gives the first hint of seizing the electronic signature of the metastable state.

The light-induced metastable state may be further inferred by fluence-dependent analysis on amplitude of the negative transient ($|\Delta$E/E$_{min}|$) and the long-lived offset ($|\Delta$E/E$_{meta}|$), respectively shown in Figure \ref{fig2} (b-c). The former increases monotonically with fluence and peculiarly becomes linear above a threshold of 56 $\mu$J/cm$^2$ (Fig. \ref{fig2}b), rather than follows a saturation behavior widely observed in topological semimetals \cite{lu2018terahertz27,george2008ultrafast28}, semiconductors and other 2D materials in which novel phase transition dynamics is absent. The latter, at the same threshold fluence, clearly shows a kink (Fig. \ref{fig2}c) and starts picking up a new component, also scaling linearly with fluence, deviating from the near-constant and negligible offset by laser heating effect. The concomitance of the threshold behavior in two qualitatively different parameters further corroborates a light-induced phase transition, for anomalies in the fluence dependence typically mean the minimal energy density required to overcome the energy barrier between two competing phases \cite{zhang2016cooperative30,stojchevska2014ultrafast31}. More strikingly, an unusual dynamic isosbestic point emerges at a fixed delay time of 3 ps from the group of pump-probe traces at above-threshold excitation (Fig. \ref{fig2}a inset). To the best of our knowledge, this is the first observation of ultrafast dynamic isosbestic point, to which a universal scaling is associated. Taken together, the abovementioned dynamic features are apparently beyond intelligible by a scenario of conventional injection and cooling of photo-doped hot carriers, but rather, indicative of percolative dynamics in the light-induced phase transition to be explicated in the end.

\begin{figure}
\includegraphics{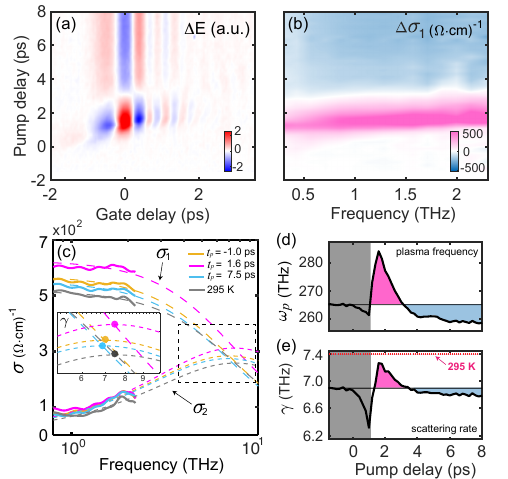}
\caption{
(a) Two-dimensional differential THz waveform as function of gate (\textit{t$_g$}) and pump delay (\textit{t$_p$}) (excitation fluence, 504 $\mu$J/cm$^2$). (b) Dynamics of the real part of the frequency-dependent THz conductivity. (c) Real and imaginary THz conductivity at selected pump delay (80 K) in comparison with that in the equilibrium state at room temperature. (experiment, solid lines; Drude-model fit, dash lines). The inset indicates scattering rates with dots. (d-e) Time-dependent plasma frequency ($\omega_p$) and carrier scattering rate ($\gamma$). The arrow indicates the equilibrium scattering rate at 295 K.
}
\label{fig3}
\end{figure}

\begin{figure*}
\includegraphics{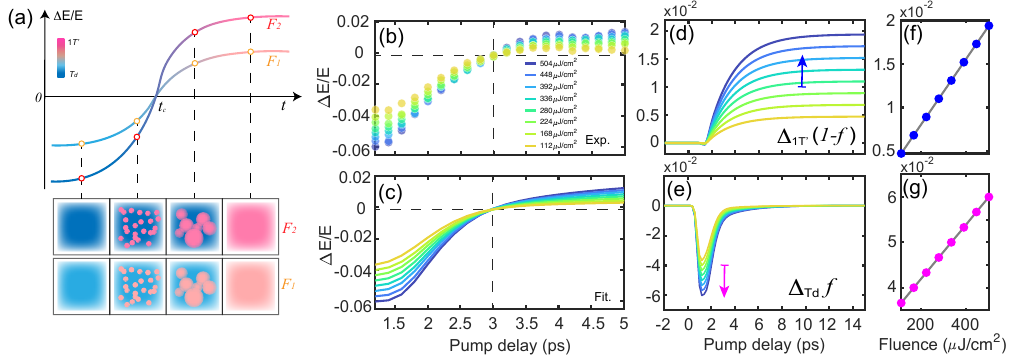}
\caption{
(a) Illustration of the percolative dynamics at high and low fluences (\textit{F$_1$}$<$\textit{F$_2$}, blue color, the photoexcited \textit{T$_d$} phase and red color the metastable 1\textit{T'} phase) (b-c) The close-up of $\Delta$E/E dynamics near the temporal isosbestic point from experiment (b) and fit to model (c). (d-e) The simulated contribution to $\Delta$E/E signal by individual percolating phases at above-threshold fluences. (f-g) Fluence dependence of the full amplitudes of the negative transient $\Delta_{1T'}$ and long-lived positive offset $\Delta_{Td}$.
}
\label{fig4}
\end{figure*}

For detailed spectroscopic information on the as-yet-purported state of electronic metastability, we now extend to a two-dimensional measurement by recording the full waveform of THz pulse, rather than just the peak amplitude, at each pump-probe delay \textit{t$_p$}. The variable relevant to the new dimension is known as the probe-gate delay (\textit{t$_g$}), with respect to which the Fourier transform gives complex transmission spectra, equivalently spectroscopic conductivity. In turn, it enables a more complete report on evolution of the free-carrier characteristics. As shown in Fig. \ref{fig3}(a-b), at a pump delay (\textit{t$_p$}) of 3 ps, the differential signals reaffirm the above isosbestic point with a complete $\pi$-phase shift in THz waveform $\Delta$E(\textit{t$_g$},\textit{t$_p$}) and a sign change in real-part conductivity $\sigma_1(\omega)$ over a spectral range of 0.2-2.5 THz. The data shows again the same transition between states dominated by photo-induced absorption ($\Delta\sigma_1$ $>$ 0) and by bleaching effect ($\Delta\sigma_1$ $<$ 0), as well as its saturation to a long-lived finite offset. 

Generally, a dynamic decrease in real-part THz conductivity ($\Delta\sigma_1$ $<$ 0) can result from the trivial thermal effect due to the increased scattering rate \cite{frenzel2013observation32,tielrooij2013photoexcitation33}, the formation of quasiparticles with enhanced effective mass \cite{suo2021observation34,lui2014trion35} or the emergence of a nontrivial metastable state accompanied by a qualitative change in electronic structure. These effects are of diametrically different origins and can be distinguished by cautious analysis of the complex THz conductivity as a whole. Fig. \ref{fig3}(c) shows spectroscopic THz conductivity at three representative pump delays (\textit{t$_p$} $=$ -1.0, 1.6 and 7.5 ps), corresponding to samples in static, transient excited and metastable states, and at the room temperature. The excellent consistency between the data and a standard Drude model ($\sigma=\omega_p^2\frac{1}{\gamma-i\omega}$) confirms the dominance of mobile carriers over low-energy electrodynamics. Trivial thermal effect can be ruled out by the fact that the free-carrier scattering in the quasi-equilibrium state (\textit{t$_p$} = 7.5 ps) is not only less than that at 295 K but also that in the equilibrium at 80 K, a trend in immediate contradiction to a laser heating scenario (inset of Fig. \ref{fig3}c). Nevertheless, the long-lived state can naturally fit in the scenario of light-induced transition into the metastable 1\textit{T'} phase, in which the sudden change in lattice symmetry removes the protection on Weyl nodes and partially drains DOS near the Fermi level. It leads to a decrease in intra-band transition amplitude, i.e., a weaker metal due to the loss in spectral weight, experimentally registered as a positive offset corresponding to an increase in THz transmission. Consistently, as shown in Fig. \ref{fig3}(d-e), the plasma frequency $\omega_p$ and scattering rate $\gamma$ show a consistent trend by surging to a peak value through electron thermalization and then decays to a long-lived offset, representing the metastable state, through electron-phonon interaction.

As for the driving mechanism of the phase transition, we find the shear phonon mode (0.24 THz) reported by other structure-sensitive techniques \cite{hein2020mode18,sie2019ultrafast19} irrelevant, owing to the absence of coherent oscillation at finite frequency resolvable by our electronic sensitive THz spectroscopy, making the quasi-DC shear displacement a more favorable candidate. To be specific, after the rapid cooling dynamics, an excessive number of electrons and holes accumulate at the band edge but can remain up to nanoseconds due to their mismatched momenta, prohibiting immediate recombination \cite{dai2015ultrafast45, wang2018room42}. As such, a favorable condition in stabilizing a particular lattice structure can be created in WTe$_2$, where occupancy near the Fermi level is decisive to configuration of the bonding orbitals in tungsten (W) and tellurium (Te) atoms. The updated interlayer interaction in the nonequilibrium state is now at play and can lead to symmetry-altering shear displacement to reach a new state of minimal energy (1\textit{T'}) \cite{kim2017origins14}. The linearity in the fluence dependence of the $\Delta$E/E offset, therefore, may indicate the magnitude of loss in DOS and even suggest a series of intermediate metastable phases accessible by controlling the key parameter, shear displacement magnitude along the \textit{T$_d$} - 1\textit{T'} transition path way.

Last but not least, we turn to the interpretation of the unusual dynamic isosbestic point shown in the above fluence-dependent study. We argue that it arises from the laser-induced percolative dynamics involving the two topologically distinct \textit{T$_d$} and 1\textit{T'} phases, schematically shown in Fig. \ref{fig4}(a). The sample starts with a \textit{T$_d$}-phase-dominated uniform state but populated with photoexcited carriers. After the further carrier proliferation at band edges by rapid thermalization dynamics, the percolation takes effect featuring a mesoscopic-scale retreating \textit{T$_d$} phase and expanding metastable 1\textit{T'} phase. In such a dynamic phase separation scenario, we assign time-dependent volume fractions $f$ and $1-f$ respectively for \textit{T$_d$} and 1\textit{T'} phases\cite{hilton2007enhanced43}, with the $f(t_p) = exp(-\frac{t_p-t_0}{\tau_0})$ describing an exponentially decaying volume fraction, where $\tau_0$ characterizes the percolation speed and $t_0$ the retarded start time of percolation owing to the finite time necessary for the photoexcited carriers to build up. Phenomenologically, each phase contributes a characteristic dynamic response in its own manner to the total $\Delta$E/E or $\Delta\sigma_1$ signals. Therefore, the $\Delta$E/E signal is a weighted average of contribution by the two competing phases, i.e., $\Delta$E/E(t)$=\Delta_{T_d}(t) \cdot f(t) + \Delta_{1T'}(t) \cdot [1-f(t)]$, where $\Delta_{T_d}(t)$ represents the dynamic negative component from a uniform \textit{T$_d$} phase with photo-injected carriers and $\Delta_{1T'}(t)$ the positive component from a uniform metastable 1\textit{T'} phase of reduced spectral weight. As shown in Fig. \ref{fig4}(c), the model features remarkable consistency with experiment and nicely repeat the dynamic isosbestic point, representing the critical timing when the competing contributions in Fig. \ref{fig4}(d-e) perfectly cancel out. Note that a success in repeating the isosbestic point implicitly requires the amplitude of percolative dynamics of both phases to covary, satisfied in WTe$_2$ by following anomalous linearity in transient peak amplitude of \textit{T$_d$}-phase character and the positive offset of 1\textit{T'}-phase character (Fig. \ref{fig4}f-g), echoing the experimental findings in Fig. \ref{fig2}(b-c). Owing to the aforementioned linearity, the timing of zero signal appears to be irrelevant to a varying excitation fluence but rather dictated by the volume fraction dynamics that embodies the intrinsic pattern of the percolation transition.

According to the general percolation theory, a percolative transition would normally display universality of some sort \cite{stanley1999scaling44}. When the tuning parameter approaches a critical value in the diabatic limit, the universality is most pronounced, and a scaling law takes effect. However, it is uncertain what the dynamic manifestation of universality would be during a phase transition in the adiabatic limit at the fundamental timescale, where classical thermodynamics may not be applicable. In Weyl semimetal WTe$_2$, the universality turns out to be a fluence-independent dynamic exchange of volume fraction between the two competing phases; that is, a universal timing of reaching a precise volume fraction for a perfect cancellation in pump-probe signals (observables) from two components of opposite sign. Regarding the macroscopic optical response, it eventuates in the remarkable dynamic isosbestic point in our far-field measurements.

To summarize, our study focuses on the ultrafast THz dynamics of a photoexcited \textit{T$_d$}-WTe$_2$ thin film. We have observed an anomalous metastable electronic state with negative THz photoconductivity. This state exhibits unique dynamical features and apparent threshold behavior. Our systematic electrodynamics analysis, combined with DFT calculation, is suggestive of reduced DOS at the Fermi level as the cause of the negative THz photoconductivity. This reduction is brought about by the light-induced \textit{T$_d$} - 1\textit{T'} structural transition of impact on topological aspects, mutually corroborated with previous studies. A phenomenological model is used to explain the unique dynamic isosbestic point, being the dynamic hallmark of the percolative phase transition in the adiabatic limit. The mechanism prompting a percolation rather than a homogenous phase transition still remains unclear and worths the attention of probes with fine spatial resolution, despite the immediate relevance to the ultrafast structural dynamics. The observation of non-equilibrium percolative dynamics at the fundamental timescale of electronic and vibrational interactions may inspire more dynamic investigation of universal behavior in light-induced phase transition in diverse systems beyond topological materials.

This research was supported by National Key Research and Development Program of China (2020YFA0309603); the Hong Kong Research Grants Council under Project No. ECS26302219, GRF16303721, GRF16306522; the National Natural Science Foundation of China Excellent Young Scientist Scheme under Grant No.12122416.

%


\end{document}